\documentclass[aps,prl,twocolumn,superscriptaddress]{revtex4-1}%
\usepackage{epsf,epsfig}
\usepackage[utf8]{inputenc}
\usepackage{amssymb,amsmath,amsfonts}
\usepackage{color}
\usepackage{slashed}
\usepackage{tensor}
\usepackage{braket}
\usepackage[colorlinks=true]{hyperref}  
\hypersetup{
    bookmarks=true,         
    unicode=false,          
    pdftoolbar=true,        
    pdfmenubar=true,        
    pdffitwindow=false,     
    pdfstartview={FitH},    
    colorlinks=true,       
    linkcolor=magenta, 
    citecolor=blue,        
    filecolor=magenta,      
    urlcolor=cyan           
} 

\newcommand{\bea}{\begin{eqnarray}}
\newcommand{\eea}{\end{eqnarray}}
\newcommand{\ba}{\begin{eqnarray}}
\newcommand{\ea}{\end{eqnarray}}

\newcommand{\beq}{\begin{equation}}
\newcommand{\eeq}{\end{equation}}
\newcommand{\beqa}{\begin{eqnarray}}
\newcommand{\eeqa}{\end{eqnarray}}
\newcommand{\beqar}{\begin{eqnarray*}}
	\newcommand{\eeqar}{\end{eqnarray*}}

\usepackage{color}

\newcommand{\req}[1]{(\ref{#1})} 

\begin{document}
\title{Resolution of Reissner-Nordstr\"om singularities by higher-derivative corrections }
\author{Pablo A. Cano}
\email{pabloantonio.cano@kuleuven.be}
\affiliation{Instituut voor Theoretische Fysica, KU Leuven, Celestijnenlaan 200D, B-3001 Leuven, Belgium}

\author{\'Angel Murcia}
\email{angel.murcia@csic.es}
\affiliation{Instituto de F\'isica Te\'orica UAM/CSIC, C/ Nicol\'as Cabrera,13-15, C.U. Cantoblanco, 28049 Madrid, Spain}

\date{June 26, 2020}

\begin{abstract}

We describe a non-minimal higher-derivative extension of Einstein-Maxwell theory in which electrically-charged black holes and point charges have globally regular gravitational and electromagnetic fields. We provide an exact static, spherically symmetric solution of this theory that reduces to the Reissner-Nordstr\"om one at weak coupling, but in which the singularity at $r=0$ is regularized for arbitrary mass and (non-vanishing) charge. We discuss the properties of these solutions and comment on the physical significance of our results.

\end{abstract}
\maketitle

One of the most anticipated features of a putative theory of Quantum Gravity is its ability to resolve the singularities that arise in General Relativity (GR). While, from a fundamental perspective, such goal is still far from being achieved, the aim of understanding this aspect of a UV-complete theory has motivated numerous research on a more phenomenological ground.
If one accepts the premise that Nature should not have singularities, then one is led to the conclusion that GR should be modified when the spacetime's curvature is extremely large. In the case of black holes, these modifications would give rise to \emph{regular} black holes, whose interior contains a smooth region of spacetime instead of a curvature singularity. 
Thus, there has been a keen interest in understanding the properties of these hypothetical regular black holes \cite{bardeen1968non,Dymnikova:1992ux,Borde:1996df,Hayward:2005gi,Frolov:2014jva,Frolov:2016pav,Carballo-Rubio:2018pmi,Simpson:2018tsi,Simpson:2019mud}, although most of the literature so far has only focused on modelizing these geometries without worrying too much about their dynamics.

When it comes to describing a dynamical foundation for regular black holes, things are much more involved.  Ideally, one would wish to find an effective high-energy modification of General Relativity whose black hole solutions were naturally singularity-free. 
There have been some interesting attempts toward this goal in the literature \cite{AyonBeato:1998ub,AyonBeato:1999ec,AyonBeato:1999rg,Balakin:2007am,Lemos:2011dq,Biswas:2011ar,Olmo:2012nx,Balakin:2015gpq,Olmo:2015axa,Fan:2016hvf,Chamseddine:2016ktu,Menchon:2017qed,Sert:2015ykz,Buoninfante:2018xiw,Buoninfante:2018stt,Giacchini:2018gxp,Cano:2018aod}, but all of them have certain limitations. For instance, some models require an unreasonable amount of fine-tuning \cite{AyonBeato:1998ub,AyonBeato:1999ec,AyonBeato:1999rg,Sert:2015ykz}, while others rely on the introduction of \emph{ad hoc} matter so that they do not possess a fully dynamical description, \textit{e.g.} \cite{Lemos:2011dq,Olmo:2015axa,Menchon:2017qed}.
On the other hand, Refs.~\cite{Biswas:2011ar,Olmo:2012nx,Buoninfante:2018xiw,Buoninfante:2018stt,Chamseddine:2016ktu} offer more promising approaches, although obtaining exact solutions is usually challenging in those cases. Despite these efforts, it is fair to say that there is still no clear way of removing curvature singularities within the framework of effective high-energy modifications of GR.

The goal of this letter is to show, in a very explicit way, that these singularities can be  indeed regularized by higher-derivative corrections.  These corrections provide a very natural extension of GR as they are generically expected to appear in the effective action of gravity. 
In particular, we are going to introduce a non-minimally coupled higher-derivative extension of Einstein-Maxwell theory that achieves the goal of regularizing physically relevant solutions of GR, namely, the electrically-charged Reissner-Nordstr\"om (RN) solution.
Besides, as we show, not only the geometry is regular in our solutions, but also the electromagnetic field remains finite everywhere. To the best of our knowledge, this is the first explicit example of a theory that fully regularizes both gravitational and electromagnetic fields for solutions of arbitrary mass and charge \footnote{In the case of magnetically-charged solutions, regular black holes were studied in Ref.~\cite{Balakin:2015gpq}  in a similar set-up. The magnetic field has a singularity though. On the other hand, in the solutions of Refs.~\cite{AyonBeato:1998ub,AyonBeato:1999ec,AyonBeato:1999rg} the charge and mass are not free parameters.}.

Without further delay, let us introduce our theory. 
We consider a non-minimally coupled action of the form
\begin{equation}\label{eq:action}
S=\frac{1}{16\pi G}\int \mathrm{d}^4x\sqrt{|g|}\left\{R-F_{\mu\nu}F_{\rho\sigma}\tensor{\chi}{^{\mu\nu\rho\sigma}}\right\}\, ,
\end{equation}
where $F_{\mu\nu}=2\partial_{[\mu}A_{\nu]}$ is the field strength of the vector field $A_{\mu}$,  $R$ is the Ricci scalar of the metric $g_{\mu\nu}$ and $\tensor{\chi}{^{\mu\nu\rho\sigma}}$ is a certain tensor built out of the curvature and the metric. Notice that when $\tensor{\chi}{^{\mu\nu}_{\rho\sigma}}=\tensor{\delta}{^{\mu\nu}_{\rho\sigma}}$ \footnote{The generalized Kronecker delta is defined as $\tensor{\delta}{^{\mu\nu}_{\rho\sigma}}=\tensor{\delta}{^{[\mu}_{[\rho}}\tensor{\delta}{^{\nu]}_{\sigma]}}$ } one gets Einstein-Maxwell theory. Instead, we are going to choose this tensor in the following way
\begin{equation}
\tensor{\chi}{^{\mu\nu}_{\rho\sigma}}=6\tensor{\delta}{^{[\mu\nu}_{\rho\sigma}}\tensor{\left(\mathcal{Q}^{-1}\right)}{^{\alpha\beta]}_{\alpha\beta}}\, ,
\end{equation}
where 
\begin{equation}
\begin{split}
\tensor{\mathcal{Q}}{^{\mu\nu}_{\rho\sigma}}&=\tensor{\delta}{^{\mu\nu}_{\rho\sigma}}+\alpha \left (6 \tensor{R}{^{[\mu}_{[\sigma}}\tensor{\delta}{^{\nu] }_{\rho]}}+7 \tensor{R}{^{\mu\nu}_{\rho\sigma}}+\frac{1}{2} R\tensor{\delta}{^{\mu\nu}_{\rho\sigma}} \right )\\&+\alpha^2 \left (\frac{9}{4} \tensor{R}{_\alpha^{[\mu}} \tensor{R}{^{\nu]\alpha}_{\rho \sigma}}+\frac{9}{4} \tensor{R}{^{\alpha}_{[\rho}} \tensor{R}{^{\mu \nu}_{\sigma] \alpha}}+\frac{1}{4} R \tensor{R}{^{\mu \nu}_{\rho \sigma}} \right. \\&\left. +\frac{35}{8} \tensor{R}{^{\mu \nu \alpha \beta}} \tensor{R}{_{ \alpha \beta \rho \sigma}}+ \frac{1}{2} \tensor{R}{_\lambda^{[\mu}} \tensor{\delta}{^{\nu] \lambda}_{\beta [\rho}} \tensor{R}{_{\sigma]}^\beta}  \right) \, ,
\end{split}
\end{equation}
and $\mathcal{Q}^{-1}$ denotes the inverse of this tensor, defined by
\begin{equation}
\tensor{\mathcal{Q}}{^{\mu\nu}_{\rho\sigma}}\tensor{\left(\mathcal{Q}^{-1}\right)}{^{\rho\sigma}_{\alpha\beta}}=\tensor{\delta}{^{\mu\nu}_{\alpha\beta}}\, .
\end{equation}
In addition, $\alpha$ is a constant with units of length squared. 
Certainly, the Lagrangian of this theory has been chosen in a very peculiar way and we will comment on this later. 
The fact that we wish to remark at this point is that, despite being fine-tuned, this theory satisfies some reasonable properties. On the one hand, it contains a free coupling constant and when it is set to zero one recovers Einstein-Maxwell theory. In fact, at low energies we have  
\begin{align}
\notag
S=&\frac{1}{16\pi G}\int \mathrm{d}^4x\sqrt{|g|}\bigg\{R-F^2+\alpha\bigg(7F_{\mu\nu}F_{\rho\sigma}\tensor{R}{^{\mu\nu\rho\sigma}}\\
&-22 F_{\mu\alpha}\tensor{F}{_{\nu}^{\alpha}}R^{\mu\nu}+\frac{9}{2}F^2R\bigg)+\mathcal{O}(\alpha^2)\bigg\}\, ,
\end{align}
so that the action reduces to the Einstein-Maxwell one when the curvature is small enough (or when $\alpha\rightarrow 0$). Thus, we can think of this theory as a toy model for a UV-completion of GR containing an infinite tower of higher-derivative terms. 
On the other hand, as we show below, for any value of the coupling we are able to find exact solutions of arbitrary mass and charge (we do not need to tune $M$ and $Q$ to particular values), making this a very useful theory for practical purposes. 

The equations of motion obtained from varying \req{eq:action} can be written in a suitable way in terms of the auxiliary tensor $\tensor{\hat{F}}{^{\mu \nu}}=\tensor{\chi}{^{\mu\nu \rho \sigma}}\tensor{F}{_{\rho \sigma}}$ as follows
\begin{align}
2\mathcal{E}^{E}_{\mu \nu}&=R_{\mu \nu}-\frac{1}{2} g_{\mu \nu} R-12 \tensor{\hat{F}}{_{\mu}^{\alpha}}\tensor{\hat{F}}{_{[\nu\alpha}}\tensor{\mathcal{Q}}{_{\rho\sigma]}^{\rho\sigma}}\nonumber  \\& +3\, g_{\mu \nu} \hat{F}^{\alpha \beta} \tensor{\hat{F}}{_{[\alpha \beta}}\tensor{\mathcal{Q}}{_{\rho\sigma]}^{\rho\sigma}} + 6\hat{F}^{\alpha \beta} \hat{F}_{[\alpha \beta} \frac{\partial\tensor{\mathcal{Q}}{_{\rho\sigma]}^{\rho\sigma}}}{\partial \tensor{R}{^{\mu\lambda\tau \gamma}}} \tensor{R}{_{\nu}^{\lambda\tau \gamma}} \nonumber\\&\label{eq:eins}+12\nabla^{\lambda} \nabla^{\gamma} \left ( \hat{F}^{\alpha \beta} \hat{F}_{[\alpha \beta} \frac{\partial\tensor{\mathcal{Q}}{_{\rho\sigma]}^{\rho\sigma}}}{\partial \tensor{R}{^{\mu \lambda \nu\gamma}}}\right )+(\mu\leftrightarrow\nu)\, , \\ 
\mathcal{E}^{M}_\nu &=\nabla_\mu\tensor{ \hat{F}}{^{\mu}_{\nu}}\,. \label{eq:max}
\end{align}
On the other hand, $F$ can be obtained from $\hat F$ using \footnote{This follows from the fact that $\tensor{(\chi^{-1})}{^{\mu\nu}_{\rho \sigma}}=6\tensor{\delta}{^{[\mu\nu}_{\rho\sigma}}\tensor{\mathcal{Q}}{^{\alpha\beta]}_{\alpha\beta}}\,$}
\begin{equation}
\tensor{F}{_{\mu \nu}}=6\tensor{\hat{F}}{_{[\rho \sigma}}\tensor{\mathcal{Q}}{_{\mu \nu]}^{\rho \sigma}} \, .
\label{eq:fandhatf}
\end{equation}
Thus, we realize that the equations of motion can be expressed solely in terms of $\hat{F}$ and $\mathcal{Q}$, hence evading the highly intricate task of computing the inverse tensor $\mathcal{Q}^{-1}$.

We shall concentrate our efforts on studying electrically-charged, static spherically symmetric solutions in this theory. 
A general ansatz for the metric and for the vector field is given by
\begin{align}
\label{eq:sssansatz}
\mathrm{d}s^2=&-N(r)^2f(r)\mathrm{d}t^2+\frac{\mathrm{d}r^2}{f(r)}+r^2\mathrm{d}\Omega_{(2)}^2\, ,\\
A=&\,\Phi(r) \mathrm{d}t\, .
\label{eq:sssansatzA}
\end{align}
The equations satisfied by the functions $N(r)$, $f(r)$ and $\Phi(r)$ are found straightforwardly by evaluating Eqs.~\eqref{eq:eins}, \eqref{eq:max} and \eqref{eq:fandhatf} above on this ansatz. In order to find a solution, we first obtain $\hat{F}$ from the Maxwell equation \eqref{eq:max}; then we substitute on the Einstein equation \eqref{eq:eins} to get the corresponding metric and finally, using \eqref{eq:fandhatf}, we obtain the field strength $F$ from where we can get the electrostatic potential $\Phi$. 
For the sake of completeness, we provide an \hyperref[app]{appendix} with the explicit form of the equations of motion and their resolution.  
\begin{figure}[t!]
\centering
\includegraphics[width=\columnwidth]{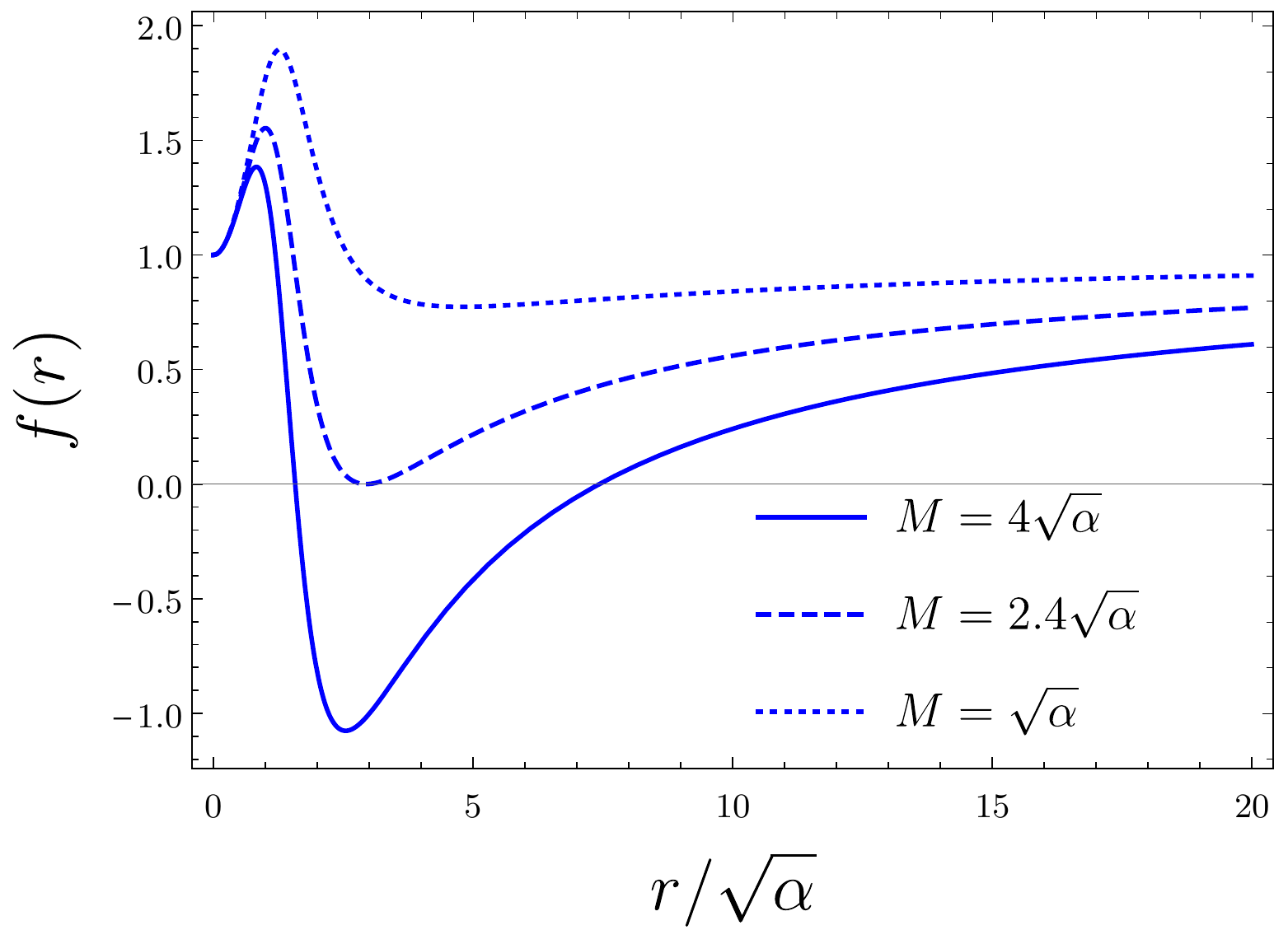}
\includegraphics[width=\columnwidth]{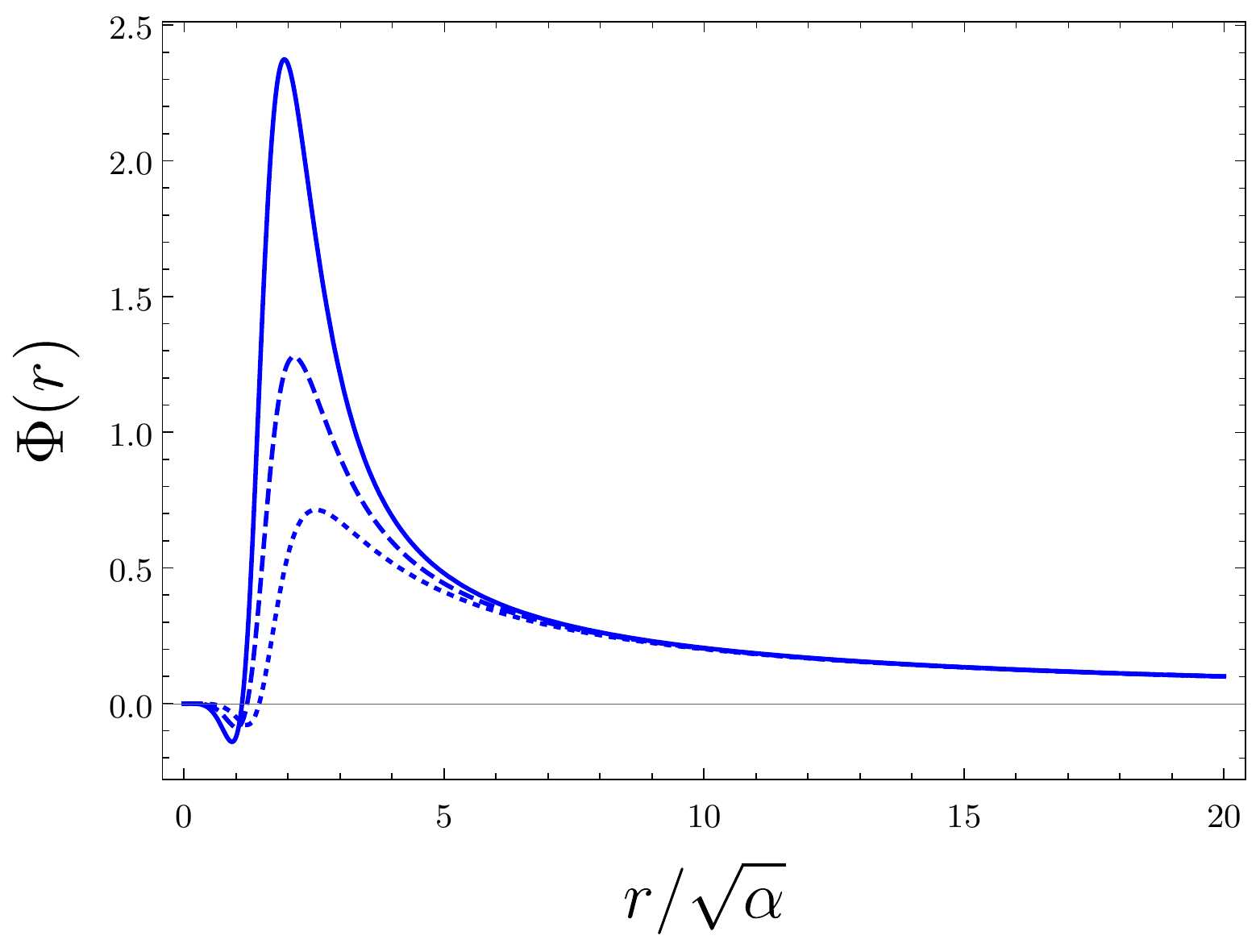}
\caption{Profile of the metric function $f(r)$ \req{eq:fsol} and the electrostatic potential $\Phi(r)$ \req{eq:phisol} as functions of the radial coordinate. In these plots we use $Q=2\sqrt{\alpha}$ and various values of the mass. Solid line: black hole with outer and inner horizons. Dashed line: extremal black hole, which in this case takes place for $M_{\mathrm{ext}} \simeq 2.4 \sqrt{\alpha}$. Dotted line: horizonless solution. In all cases the point $r=0$ is a smooth cap of the geometry and both $f(r)$ and $\Phi(r)$ are finite everywhere.} 
\label{fig:sol}
\end{figure}

Remarkably, we find that the equations can be fully integrated, and the most general asymptotically flat solution reads
\begin{align}
N(r)=&\, 1\,,\\
\label{eq:fsol}
f(r)=&\, \frac{r^4 \left(r^2-2 M r+Q^2\right)+\alpha  Q^2 (3 r^2+2 \alpha)}{r^6+\alpha  Q^2 (r^2+2 \alpha)}\, ,\\
\Phi(r)=&\frac{Q}{r} \left(1+\frac{\alpha(1-f)}{r^2}\right) \left(1+\frac{\alpha(4 -4 f+ r f')}{2r^2}\right)\, ,
\label{eq:phisol}
\end{align}
where $M$ and $Q$ are two integration constants. 
The profile of the functions $f(r)$ and $\Phi(r)$ is shown in Fig.~\ref{fig:sol} for specific values of these parameters.

Let us now examine the properties of this solution. First, notice that asymptotically the fields behave as in the Reissner-Nordstr\"om  solution, $f(r)\sim 1-2M/r+Q^2/r^2$, $\Phi(r)\sim Q/r$, from where one can identify $M$ with the mass and $Q$ with the electric charge. Also, the full RN solution is recovered when we set $\alpha=0$. Thus, this solution is a continuous deformation of the RN one and the deviations with respect to it are small as long as the curvature and the field strength take sufficiently small values. On the other hand, the corrections have a drastic effect near the would-be singularities, where these quantities would diverge. 
Indeed, the most remarkable property of this solution, as can be easily seen from \req{eq:fsol}, is that the geometry is smooth everywhere. More precisely, we check that $f(r)$ has no divergences, and we find that around $r=0$  it behaves as
\begin{equation}
f(r)=1+\frac{r^2}{\alpha}+\mathcal{O}(r^3)\, ,
\end{equation}
which implies that the point $r=0$ is a smooth cap of the geometry. In particular, the region near $r=0$ is locally that of AdS space of radius $\sqrt{\alpha}$. Interestingly enough, the electromagnetic field is also finite everywhere and one can check that near the origin the electrostatic potential is given by
\begin{equation}
\Phi(r)\sim -\frac{M^2 r^5}{2 Q^3\alpha^2}+\mathcal{O}(r^7)\, .
\end{equation}
The field strength $F=-\Phi' \mathrm{d}t\wedge \mathrm{d}r$ is also finite and vanishes at $r=0$. Thus, these solutions are free of singularities.

Depending on the relative values of the mass and the charge, these solutions have a different nature.
As we can see in Eq.~\req{eq:fsol}, only the term proportional to the mass comes with a negative sign, which means that, if $M$ is  large enough compared to the charge, $f(r)$ will vanish at certain point. In that case the solution contains a horizon and hence it is a black hole --- see the solid lines in Fig.~\ref{fig:sol}. In addition, this black hole has always a second, inner horizon (except in the extremal limit), so the causal structure is very similar to that of the RN black hole, with the difference that the timelike singularity at $r=0$ is removed. The Penrose diagram of this regular black hole is identical to that of other models often discussed in the literature \cite{Hayward:2005gi,Frolov:2016pav}, so we refer to those works for further details. 
For a specific value of the mass, $M=M_{\rm ext}(Q)$, both horizons merge into a degenerate horizon and we have a extremal black hole --- depicted by the dashed lines in Fig.~\ref{fig:sol}. Let us note that the extremality condition is modified with respect to the case of the Reissner-Nordstr\"om black hole, so that $M_{\rm ext}\neq Q$.
Finally, if the mass is below the extremal value, then the effect of the charge dominates and the solution does not possess a horizon. This situation is represented by the dotted lines in Fig.~\ref{fig:sol}. This horizonless smooth solution is particularly interesting and, in principle, one could think of it as a soliton or a fuzzball.

\begin{figure}[t]
\centering
\includegraphics[width=\columnwidth]{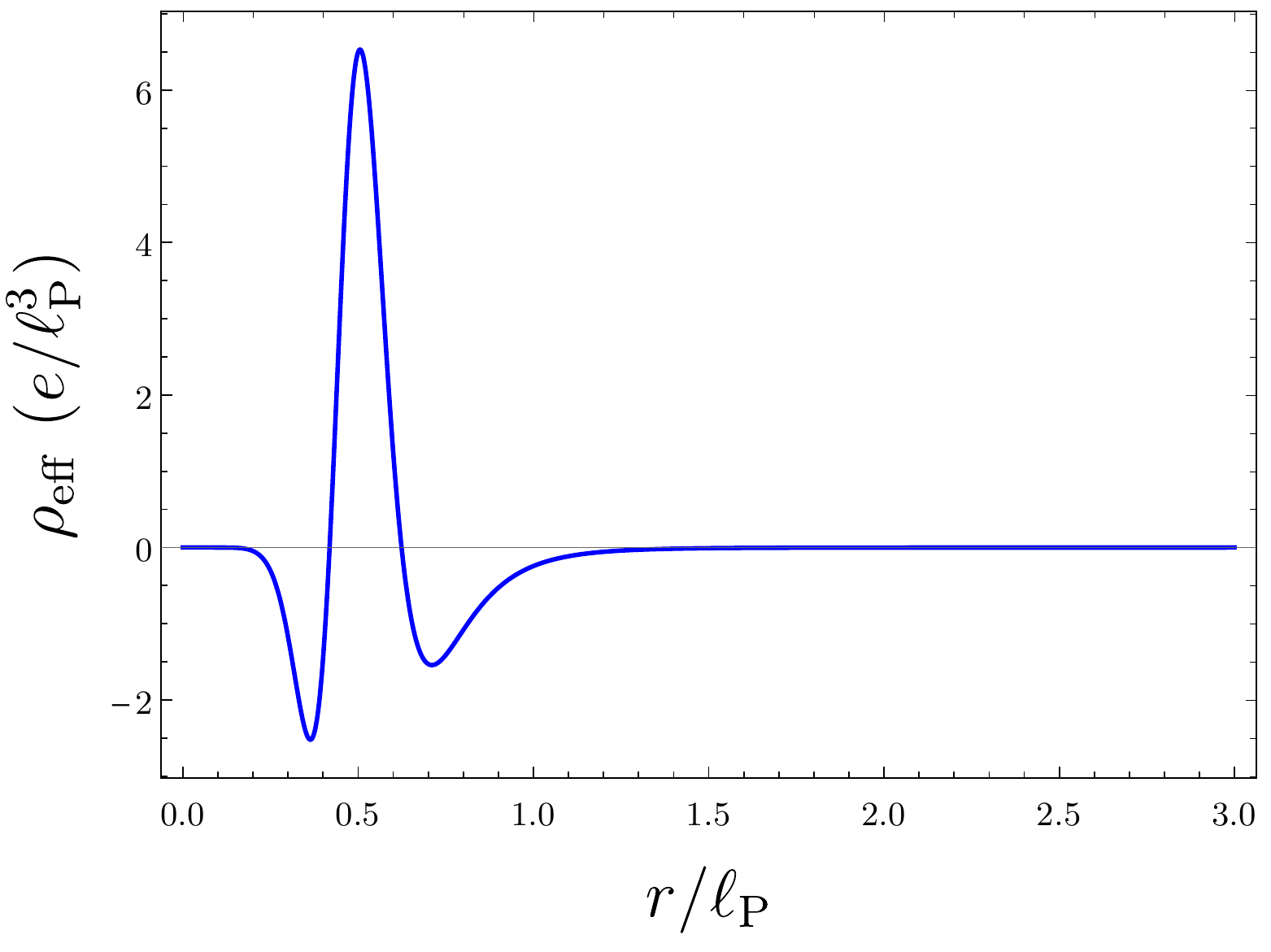}
\caption{Effective charge density of an electron (defined as in Eq.~\req{eq:rhoeff}) predicted by the theory \req{eq:action}. We assume that the corrections appear at Planck scale, $\alpha=\ell_{\rm P}^2$.} 
\label{fig:rho}
\end{figure}

One intriguing question is that about the origin of the charge and the mass in these solutions. Apparently, there are no matter sources involved, so one might conclude that the mass and charge arise due to the non-linear interactions between gravity and electromagnetism. 
However, a closer look reveals that this is not entirely correct. While the geometry is smooth ($\mathcal{C}^{\infty}$) everywhere, one can check that the potential $\Phi$ (and hence the vector $A$) is only $\mathcal{C}^{4}$ at $r=0$. This means that some of the equations of motion may not be satisfied at $r=0$, which typically indicates the presence of point-like sources.  In fact, in the case of the Maxwell equation \req{eq:max} it is easy to see that we have a Dirac delta on the right-hand-side,
\begin{equation}
\nabla_{\mu}\hat F^{\mu\nu}=4 \pi Q\delta^{(3)} (r)\delta^{\nu}_{t}\, ,
\end{equation}
and hence, these solutions do have a point-like source of electric charge.  This means that the horizonless solutions we have found should be really interpreted as the fields of charged point particles, and not as solitons. 
Thus, the higher-derivatives seem to have the effect of ``smearing'' the charge, so that, even though it is concentrated at a single point, the fields are finite. As an interesting example, we may consider the case of an electron. Let us assume that the corrections appear at Planck scale and that $\alpha=\ell_{\rm P}^2$. In Planck units the charge of the electron is $Q=-e G^{1/2}(4\pi\epsilon_0)^{-1/2}\approx -1/\sqrt{137}$, where $e$ is the elementary charge. We may also approximate $m_e\approx 0$, since it is much smaller than Planck's mass. Then, even though we treat the electron as a point particle, we may define an effective charge density in the usual way, $\mathrm{d}\star F=4 \pi \rho_{\rm eff} V_3$, where $V_3$ is the volume form of constant-$t$ spatial slices. This leads to the identification
\begin{equation}\label{eq:rhoeff}
4 \pi \rho_{\rm eff}=-\frac{\sqrt{f}}{r^2}\frac{d}{dr}\left(r^2\frac{d\Phi}{dr}\right)\, .
\end{equation}
The profile of this effective charge density is shown in Fig.~\ref{fig:rho}, where we can check that it is finite everywhere and concentrated around the region $r<\ell_{\rm P}$. Thus, it is as if the higher-derivative corrections delocalize the charge, yielding the electron some apparent structure, which in this case shows up at Planck scale. On the other hand, the gravitational and electromagnetic potentials, $f(r)$ and $\Phi(r)$ respectively, have qualitatively similar profiles to those shown in the dotted lines in Fig.~\ref{fig:sol}. All of this provides us with a remarkable physical picture. In the first place, the electron sources the electromagnetic field, which at the same time creates a gravitational field. Then, due to the non-minimal couplings between them, a non-linear backreaction is produced which at the end renders both fields finite.

Summarizing, we have shown that the theory \req{eq:action} is able to resolve the singularities of charged black holes and of point-like charge particles. The regular black holes we have obtained have similar properties to some of the models analyzed in the literature \cite{Hayward:2005gi,Frolov:2016pav}, with the timelike singularity of RN black holes replaced by a smooth ``AdS core''. On the other hand, point charges acquire an effective structure of finite size due to the short-distance modifications of gravity and electromagnetism implied by the theory \req{eq:action}. 
Thus, we have proven that the regularization of singularities \emph{is} possible within the framework of Einstein-Maxwell theory with higher derivatives.  Note that we do not intend to say that such regularization is a general phenomenon. It suffices to show that it can be achieved by some theories to prove that effective actions can capture this property of a UV-complete theory. 
In fact, this makes the action \req{eq:action} a very interesting model for a UV-completion of Einstein-Maxwell theory.

Let us close this letter by further commenting on this theory. The Lagrangian of \req{eq:action} has been found after imposing mainly technical conditions. Namely, it is characterized by having static spherically symmetric solutions satisfying $g_{tt}g_{rr}=-1$, and for which the equations of motion can be fully integrated. These properties resemble those of the recently constructed Generalized Quasitopological gravities, \textit{e.g.} \cite{PabloPablo,Hennigar:2016gkm,PabloPablo2,Hennigar:2017ego,PabloPablo3}, but while those theories are purely gravitational, the action \req{eq:action} can be thought as a non-minimally coupled version thereof.  Now, it turns out that there are many other non-minimal higher-derivative extensions of Einstein-Maxwell theory of this type \cite{Cano:2020qhy} --- see also \cite{MuellerHoissen:1988bp,Balakin:2016mnn}. Within this class of theories, one finds that the resolution of curvature singularities happens generally, while only a few constraints need to be imposed in order to regularize the electromagnetic field. The Lagrangian in Eq.~\req{eq:action} is a particular case of this, but many other examples can be found as well \cite{Cano:2020qhy}. Thus, most of the fine-tuning in \req{eq:action} actually comes from imposing that the equations of motion can be solved exactly, not from a physical requirement. Probably, many other higher-derivative theories are also singularity-free, but one cannot check this easily due to the complicated form of the equations of motion in the general case. 
This suggests that the regularization of singularities by non-minimal higher-derivative terms as those in \req{eq:action} could be, after all, a more general phenomenon than expected.

\vspace{0.4cm}
\begin{acknowledgments}   
\textbf{\textit{Acknowledgments.}}
We would like to thank Pablo Bueno, Pedro F. Ram\'irez and Carlos S. Shahbazi for useful comments and discussions. 
The work of PAC is supported by the C16/16/005 grant of the KU Leuven. The work of \'AM is funded by the Spanish FPU Grant No. FPU17/04964. \'AM was further supported by the MCIU/AEI/FEDER UE grant PGC2018-095205-B-I00, and by the ``Centro de Excelencia Severo Ochoa'' Program grant SEV-2016-0597.

\end{acknowledgments}

\onecolumngrid  
\appendix 

\vspace{0.5cm}
\section{\large Appendix: solving the equations of motion}
\label{app}
First, one can see that, due to the form of the ansatz in Eqs~\req{eq:sssansatz} and \req{eq:sssansatzA}, the only non-vanishing component of $\hat F_{\mu\nu}$ is $\hat F_{tr}=-\hat F_{rt}$. Then, Eq.~\eqref{eq:max} implies that this tensor must have the form
\begin{equation}
\hat{F}=\frac{Q N(r)}{r^2} \mathrm{d} t \wedge \mathrm{d} r\,,
\label{eq:hatf}
\end{equation}
where $Q$ is an integration constant that (as we check later) represents the electric charge.  The next step is to substitute \eqref{eq:hatf} on the Einstein equation \eqref{eq:eins}. Before that, we note that the static condition and spherical symmetry imply that all its off-diagonal components vanish identically. Furthermore, using the Bianchi identity one may deduce that the angular components are satisfied once the $tt$ and $rr$ components hold. Therefore these are the only non-trivial equations we obtain from \eqref{eq:eins}.
Imposing the static and spherically symmetric ansatz \eqref{eq:sssansatz} together with the expression of $\hat{F}$ presented in Eq. \eqref{eq:hatf}, we find:
\begin{align}
\notag
\mathcal{E}_{tt}^E=&-\frac{f N^2}{r^4} \left(r^3
   f'+(f-1) r^2+Q^2\right)-\frac{\alpha  f N^2 Q^2}{r^6} \left(r f'-3 f+9\right)\\
   &+\frac{\alpha ^2 f Q^2}{2 r^8} \left(-f^2 N'^2 r^2+4
   N^2 \left(-r f'+5 f-5\right)+2 f N r \left(2 N' r f'-4 f
  N'+f N'' r\right)\right)=0\, ,\label{eq:tteq}\\
\mathcal{E}_{tt}^E+f^2 N^2 \mathcal{E}_{rr}^E=&2 N N' f^2  \left( \frac{1}{r}+\frac{Q^2 \alpha}{r^5}-\frac{Q^2 \alpha^2 (-2N+rf N')}{N r^7} \right)=0\, .
\label{eq:rrtteq}
\end{align}
Despite the intricateness of Eq. \eqref{eq:tteq}, we see by direct inspection that the combination of $\mathcal{E}_{tt}^E $ and $\mathcal{E}_{rr}^E $ in Eq.~\req{eq:rrtteq} imposes $N=$ constant (another possible solution could be  obtained by setting the quantity between brackets to zero, but this yields an unphysical solution which is not asymptotically flat).  In particular, we may always set $N=1$ after an appropriate rescaling of the time coordinate. Imposing $N=1$ simplifies Eq. \eqref{eq:tteq}, which takes the form
\begin{equation}
\left. \mathcal{E}_{tt}^E \right \vert_{N=1}= -\frac{f}{r^4} \left(r^3
   f'+(f-1) r^2+Q^2\right)-\frac{\alpha  Q^2 f }{r^6}\left(r f'-3 f+9\right)-\frac{2 \alpha ^2 Q^2 f}{r^8} \left(5-5f+r f'\right)=0\,.
\end{equation}
Interestingly enough, the combination $\dfrac{\left. r^2 \mathcal{E}_{tt}^E \right \vert_{N=1} }{f}$ can be easily integrated. One finds that 
\begin{equation}
\frac{1}{2} \int \frac{\left. r^2 \mathcal{E}_{tt}^E \right \vert_{N=1} }{f} \mathrm{d}r=r(1-f)+\frac{Q^2}{r}+\frac{3 \alpha Q^2}{r^3}-\frac{ \alpha Q^2  f}{r^3}+\frac{(1-f)}{r^5} 2\alpha^2 Q^2=2M\, ,
\end{equation}
where $M$ is an integration constant that we identify with the mass of the solution. Thus, we have a linear equation for the metric function $f$ whose solution reads
\begin{equation}
f(r)= \frac{r^4 \left(r^2-2 M r+Q^2\right)+\alpha  Q^2 (3 r^2+2 \alpha)}{r^6+\alpha  Q^2 (r^2+2 \alpha)}\, .
\end{equation}
This is the expression for $f$ given at Eq. \eqref{eq:fsol}. The following task is to derive the original gauge field strength $F$. Using Eq. \eqref{eq:fandhatf} we have that
\begin{equation}
F_{\mu \nu}=\tensor{\mathcal{Q}}{_{\mu \nu}^{\rho \sigma}} \hat{F}_{\rho \sigma}+4 \tensor{\mathcal{Q}}{^{\alpha \beta}_{\alpha[\mu}} \tensor{\hat{F}}{_{\nu] \beta}}+\tensor{\mathcal{Q}}{_{\alpha \beta}^{\alpha \beta}} \hat{F}_{\mu \nu}\,.
\label{eq:eqforf}
\end{equation}
On the one hand, $\hat{F}$ was already obtained back at Eq. \eqref{eq:hatf}. On the other hand, we have just derived the expression for the metric functions $f$ and $N$, so the tensor $\tensor{\mathcal{Q}}{_{\mu \nu}^{\rho \sigma}}$ is also determined as well. Assuming the electric ansatz \eqref{eq:sssansatzA}, which implies that
\begin{equation}
F=-\Phi'(r) \mathrm{d}t \wedge \mathrm{d}r\,,
\end{equation}
we find, after equating this last expression with Eq. \eqref{eq:eqforf}, a first-order ODE for $\Phi(r)$ in terms of the electric charge $Q$ and the metric functions $f$ and $N$. Such equation reads
\begin{equation}
\begin{split}
\Phi'(r)&=-\frac{Q N}{r^2}+\frac{\alpha  Q}{2 r^4} \Big[r \left(3 r f' N'+f \left(2 r N''-8 N' \right)\right)+N \left(r^2 f''-8 r f'+18 f-18\right)\Big]\\&-\frac{\alpha ^2 Q}{2 r^5} \Big[-3 r f' N'+f \left(\left(5 r f'+12\right) N'-2 r N''\right)+2 f^2 \left(r N''-6 N'\right)\Big]\\&-\frac{\alpha^2 Q }{2 r^6 N}\Big[ N^2 \left(-r^2 f''+r^2 f'^2+12 r f'+f \left(r^2 f''-12 r f'-40\right)+20 f^2+20\right)+r^2 f^2 N'^2\Big]\, ,
\end{split}
\label{eq:odephi}
\end{equation}
where we remark that we have not replaced yet the expressions obtained for $N$ and $f$.
After setting $N=1$, however, we find that a great simplification takes place and \eqref{eq:odephi} boils down to
\begin{equation}
\Phi'(r)=-\frac{Q}{r^2}+\frac{\alpha  Q}{2r^4} \left(r^2 f''-8 r f'+18 f-18\right) -\frac{\alpha ^2 Q}{2 r^6}\Big[ \left(20-r^2 f''+r^2 f'^2+12 r f'\right )+ f \left(r^2 f''+20 f-12 r f'-40\right)\Big]\,.
\end{equation}
Imposing a vanishing electric potential at infinity, this equation can be integrated to yield
\begin{equation}
\Phi(r)=\frac{Q}{r} \left(1+\frac{\alpha(1-f)}{r^2}\right) \left(1+\frac{\alpha(4 -4 f+ r f')}{2r^2}\right)\, ,
\end{equation}
which is precisely Eq. \eqref{eq:phisol}.

\bibliographystyle{apsrev4-1} 

\bibliography{Gravities}

 \newcommand{\noop}[1]{}
\begin{thebibliography}{37}%
\makeatletter
\providecommand \@ifxundefined [1]{%
 \@ifx{#1\undefined}
}%
\providecommand \@ifnum [1]{%
 \ifnum #1\expandafter \@firstoftwo
 \else \expandafter \@secondoftwo
 \fi
}%
\providecommand \@ifx [1]{%
 \ifx #1\expandafter \@firstoftwo
 \else \expandafter \@secondoftwo
 \fi
}%
\providecommand \natexlab [1]{#1}%
\providecommand \enquote  [1]{``#1''}%
\providecommand \bibnamefont  [1]{#1}%
\providecommand \bibfnamefont [1]{#1}%
\providecommand \citenamefont [1]{#1}%
\providecommand \href@noop [0]{\@secondoftwo}%
\providecommand \href [0]{\begingroup \@sanitize@url \@href}%
\providecommand \@href[1]{\@@startlink{#1}\@@href}%
\providecommand \@@href[1]{\endgroup#1\@@endlink}%
\providecommand \@sanitize@url [0]{\catcode `\\12\catcode `\$12\catcode
  `\&12\catcode `\#12\catcode `\^12\catcode `\_12\catcode `\%12\relax}%
\providecommand \@@startlink[1]{}%
\providecommand \@@endlink[0]{}%
\providecommand \url  [0]{\begingroup\@sanitize@url \@url }%
\providecommand \@url [1]{\endgroup\@href {#1}{\urlprefix }}%
\providecommand \urlprefix  [0]{URL }%
\providecommand \Eprint [0]{\href }%
\providecommand \doibase [0]{http://dx.doi.org/}%
\providecommand \selectlanguage [0]{\@gobble}%
\providecommand \bibinfo  [0]{\@secondoftwo}%
\providecommand \bibfield  [0]{\@secondoftwo}%
\providecommand \translation [1]{[#1]}%
\providecommand \BibitemOpen [0]{}%
\providecommand \bibitemStop [0]{}%
\providecommand \bibitemNoStop [0]{.\EOS\space}%
\providecommand \EOS [0]{\spacefactor3000\relax}%
\providecommand \BibitemShut  [1]{\csname bibitem#1\endcsname}%
\let\auto@bib@innerbib\@empty
\bibitem [{\citenamefont {Bardeen}(1968)}]{bardeen1968non}%
  \BibitemOpen
  \bibfield  {author} {\bibinfo {author} {\bibfnamefont {J.~M.}\ \bibnamefont
  {Bardeen}},\ }in\ \href@noop {} {\emph {\bibinfo {booktitle} {Proc. Int.
  Conf. GR5, Tbilisi}}},\ Vol.\ \bibinfo {volume} {174}\ (\bibinfo {year}
  {1968})\BibitemShut {NoStop}%
\bibitem [{\citenamefont {Dymnikova}(1992)}]{Dymnikova:1992ux}%
  \BibitemOpen
  \bibfield  {author} {\bibinfo {author} {\bibfnamefont {I.}~\bibnamefont
  {Dymnikova}},\ }\href {\doibase 10.1007/BF00760226} {\bibfield  {journal}
  {\bibinfo  {journal} {Gen. Rel. Grav.}\ }\textbf {\bibinfo {volume} {24}},\
  \bibinfo {pages} {235} (\bibinfo {year} {1992})}\BibitemShut {NoStop}%
\bibitem [{\citenamefont {Borde}(1997)}]{Borde:1996df}%
  \BibitemOpen
  \bibfield  {author} {\bibinfo {author} {\bibfnamefont {A.}~\bibnamefont
  {Borde}},\ }\href {\doibase 10.1103/PhysRevD.55.7615} {\bibfield  {journal}
  {\bibinfo  {journal} {Phys. Rev.}\ }\textbf {\bibinfo {volume} {D55}},\
  \bibinfo {pages} {7615} (\bibinfo {year} {1997})},\ \Eprint
  {http://arxiv.org/abs/gr-qc/9612057} {arXiv:gr-qc/9612057 [gr-qc]}
  \BibitemShut {NoStop}%
\bibitem [{\citenamefont {Hayward}(2006)}]{Hayward:2005gi}%
  \BibitemOpen
  \bibfield  {author} {\bibinfo {author} {\bibfnamefont {S.~A.}\ \bibnamefont
  {Hayward}},\ }\href {\doibase 10.1103/PhysRevLett.96.031103} {\bibfield
  {journal} {\bibinfo  {journal} {Phys. Rev. Lett.}\ }\textbf {\bibinfo
  {volume} {96}},\ \bibinfo {pages} {031103} (\bibinfo {year} {2006})},\
  \Eprint {http://arxiv.org/abs/gr-qc/0506126} {arXiv:gr-qc/0506126}
  \BibitemShut {NoStop}%
\bibitem [{\citenamefont {Frolov}(2014)}]{Frolov:2014jva}%
  \BibitemOpen
  \bibfield  {author} {\bibinfo {author} {\bibfnamefont {V.~P.}\ \bibnamefont
  {Frolov}},\ }\href {\doibase 10.1007/JHEP05(2014)049} {\bibfield  {journal}
  {\bibinfo  {journal} {JHEP}\ }\textbf {\bibinfo {volume} {05}},\ \bibinfo
  {pages} {049} (\bibinfo {year} {2014})},\ \Eprint
  {http://arxiv.org/abs/1402.5446} {arXiv:1402.5446 [hep-th]} \BibitemShut
  {NoStop}%
\bibitem [{\citenamefont {Frolov}(2016)}]{Frolov:2016pav}%
  \BibitemOpen
  \bibfield  {author} {\bibinfo {author} {\bibfnamefont {V.~P.}\ \bibnamefont
  {Frolov}},\ }\href {\doibase 10.1103/PhysRevD.94.104056} {\bibfield
  {journal} {\bibinfo  {journal} {Phys. Rev.}\ }\textbf {\bibinfo {volume}
  {D94}},\ \bibinfo {pages} {104056} (\bibinfo {year} {2016})},\ \Eprint
  {http://arxiv.org/abs/1609.01758} {arXiv:1609.01758 [gr-qc]} \BibitemShut
  {NoStop}%
\bibitem [{\citenamefont {Carballo-Rubio}\ \emph {et~al.}(2018)\citenamefont
  {Carballo-Rubio}, \citenamefont {Di~Filippo}, \citenamefont {Liberati},
  \citenamefont {Pacilio},\ and\ \citenamefont
  {Visser}}]{Carballo-Rubio:2018pmi}%
  \BibitemOpen
  \bibfield  {author} {\bibinfo {author} {\bibfnamefont {R.}~\bibnamefont
  {Carballo-Rubio}}, \bibinfo {author} {\bibfnamefont {F.}~\bibnamefont
  {Di~Filippo}}, \bibinfo {author} {\bibfnamefont {S.}~\bibnamefont
  {Liberati}}, \bibinfo {author} {\bibfnamefont {C.}~\bibnamefont {Pacilio}}, \
  and\ \bibinfo {author} {\bibfnamefont {M.}~\bibnamefont {Visser}},\ }\href
  {\doibase 10.1007/JHEP07(2018)023} {\bibfield  {journal} {\bibinfo  {journal}
  {JHEP}\ }\textbf {\bibinfo {volume} {07}},\ \bibinfo {pages} {023} (\bibinfo
  {year} {2018})},\ \Eprint {http://arxiv.org/abs/1805.02675} {arXiv:1805.02675
  [gr-qc]} \BibitemShut {NoStop}%
\bibitem [{\citenamefont {Simpson}\ and\ \citenamefont
  {Visser}(2019{\natexlab{a}})}]{Simpson:2018tsi}%
  \BibitemOpen
  \bibfield  {author} {\bibinfo {author} {\bibfnamefont {A.}~\bibnamefont
  {Simpson}}\ and\ \bibinfo {author} {\bibfnamefont {M.}~\bibnamefont
  {Visser}},\ }\href {\doibase 10.1088/1475-7516/2019/02/042} {\bibfield
  {journal} {\bibinfo  {journal} {JCAP}\ }\textbf {\bibinfo {volume} {02}},\
  \bibinfo {pages} {042} (\bibinfo {year} {2019}{\natexlab{a}})},\ \Eprint
  {http://arxiv.org/abs/1812.07114} {arXiv:1812.07114 [gr-qc]} \BibitemShut
  {NoStop}%
\bibitem [{\citenamefont {Simpson}\ and\ \citenamefont
  {Visser}(2019{\natexlab{b}})}]{Simpson:2019mud}%
  \BibitemOpen
  \bibfield  {author} {\bibinfo {author} {\bibfnamefont {A.}~\bibnamefont
  {Simpson}}\ and\ \bibinfo {author} {\bibfnamefont {M.}~\bibnamefont
  {Visser}},\ }\href {\doibase 10.3390/universe6010008} {\bibfield  {journal}
  {\bibinfo  {journal} {Universe}\ }\textbf {\bibinfo {volume} {6}},\ \bibinfo
  {pages} {8} (\bibinfo {year} {2019}{\natexlab{b}})},\ \Eprint
  {http://arxiv.org/abs/1911.01020} {arXiv:1911.01020 [gr-qc]} \BibitemShut
  {NoStop}%
\bibitem [{\citenamefont {Ayon-Beato}\ and\ \citenamefont
  {Garcia}(1998)}]{AyonBeato:1998ub}%
  \BibitemOpen
  \bibfield  {author} {\bibinfo {author} {\bibfnamefont {E.}~\bibnamefont
  {Ayon-Beato}}\ and\ \bibinfo {author} {\bibfnamefont {A.}~\bibnamefont
  {Garcia}},\ }\href {\doibase 10.1103/PhysRevLett.80.5056} {\bibfield
  {journal} {\bibinfo  {journal} {Phys. Rev. Lett.}\ }\textbf {\bibinfo
  {volume} {80}},\ \bibinfo {pages} {5056} (\bibinfo {year} {1998})},\ \Eprint
  {http://arxiv.org/abs/gr-qc/9911046} {arXiv:gr-qc/9911046} \BibitemShut
  {NoStop}%
\bibitem [{\citenamefont {Ayon-Beato}\ and\ \citenamefont
  {Garcia}(1999{\natexlab{a}})}]{AyonBeato:1999ec}%
  \BibitemOpen
  \bibfield  {author} {\bibinfo {author} {\bibfnamefont {E.}~\bibnamefont
  {Ayon-Beato}}\ and\ \bibinfo {author} {\bibfnamefont {A.}~\bibnamefont
  {Garcia}},\ }\href {\doibase 10.1023/A:1026640911319} {\bibfield  {journal}
  {\bibinfo  {journal} {Gen. Rel. Grav.}\ }\textbf {\bibinfo {volume} {31}},\
  \bibinfo {pages} {629} (\bibinfo {year} {1999}{\natexlab{a}})},\ \Eprint
  {http://arxiv.org/abs/gr-qc/9911084} {arXiv:gr-qc/9911084} \BibitemShut
  {NoStop}%
\bibitem [{\citenamefont {Ayon-Beato}\ and\ \citenamefont
  {Garcia}(1999{\natexlab{b}})}]{AyonBeato:1999rg}%
  \BibitemOpen
  \bibfield  {author} {\bibinfo {author} {\bibfnamefont {E.}~\bibnamefont
  {Ayon-Beato}}\ and\ \bibinfo {author} {\bibfnamefont {A.}~\bibnamefont
  {Garcia}},\ }\href {\doibase 10.1016/S0370-2693(99)01038-2} {\bibfield
  {journal} {\bibinfo  {journal} {Phys. Lett. B}\ }\textbf {\bibinfo {volume}
  {464}},\ \bibinfo {pages} {25} (\bibinfo {year} {1999}{\natexlab{b}})},\
  \Eprint {http://arxiv.org/abs/hep-th/9911174} {arXiv:hep-th/9911174}
  \BibitemShut {NoStop}%
\bibitem [{\citenamefont {Balakin}\ \emph {et~al.}(2008)\citenamefont
  {Balakin}, \citenamefont {Bochkarev},\ and\ \citenamefont
  {Lemos}}]{Balakin:2007am}%
  \BibitemOpen
  \bibfield  {author} {\bibinfo {author} {\bibfnamefont {A.~B.}\ \bibnamefont
  {Balakin}}, \bibinfo {author} {\bibfnamefont {V.~V.}\ \bibnamefont
  {Bochkarev}}, \ and\ \bibinfo {author} {\bibfnamefont {J.~P.}\ \bibnamefont
  {Lemos}},\ }\href {\doibase 10.1103/PhysRevD.77.084013} {\bibfield  {journal}
  {\bibinfo  {journal} {Phys. Rev. D}\ }\textbf {\bibinfo {volume} {77}},\
  \bibinfo {pages} {084013} (\bibinfo {year} {2008})},\ \Eprint
  {http://arxiv.org/abs/0712.4066} {arXiv:0712.4066 [gr-qc]} \BibitemShut
  {NoStop}%
\bibitem [{\citenamefont {Lemos}\ and\ \citenamefont
  {Zanchin}(2011)}]{Lemos:2011dq}%
  \BibitemOpen
  \bibfield  {author} {\bibinfo {author} {\bibfnamefont {J.~P.}\ \bibnamefont
  {Lemos}}\ and\ \bibinfo {author} {\bibfnamefont {V.~T.}\ \bibnamefont
  {Zanchin}},\ }\href {\doibase 10.1103/PhysRevD.83.124005} {\bibfield
  {journal} {\bibinfo  {journal} {Phys. Rev. D}\ }\textbf {\bibinfo {volume}
  {83}},\ \bibinfo {pages} {124005} (\bibinfo {year} {2011})},\ \Eprint
  {http://arxiv.org/abs/1104.4790} {arXiv:1104.4790 [gr-qc]} \BibitemShut
  {NoStop}%
\bibitem [{\citenamefont {Biswas}\ \emph {et~al.}(2012)\citenamefont {Biswas},
  \citenamefont {Gerwick}, \citenamefont {Koivisto},\ and\ \citenamefont
  {Mazumdar}}]{Biswas:2011ar}%
  \BibitemOpen
  \bibfield  {author} {\bibinfo {author} {\bibfnamefont {T.}~\bibnamefont
  {Biswas}}, \bibinfo {author} {\bibfnamefont {E.}~\bibnamefont {Gerwick}},
  \bibinfo {author} {\bibfnamefont {T.}~\bibnamefont {Koivisto}}, \ and\
  \bibinfo {author} {\bibfnamefont {A.}~\bibnamefont {Mazumdar}},\ }\href
  {\doibase 10.1103/PhysRevLett.108.031101} {\bibfield  {journal} {\bibinfo
  {journal} {Phys. Rev. Lett.}\ }\textbf {\bibinfo {volume} {108}},\ \bibinfo
  {pages} {031101} (\bibinfo {year} {2012})},\ \Eprint
  {http://arxiv.org/abs/1110.5249} {arXiv:1110.5249 [gr-qc]} \BibitemShut
  {NoStop}%
\bibitem [{\citenamefont {Olmo}\ and\ \citenamefont
  {Rubiera-Garcia}(2012)}]{Olmo:2012nx}%
  \BibitemOpen
  \bibfield  {author} {\bibinfo {author} {\bibfnamefont {G.~J.}\ \bibnamefont
  {Olmo}}\ and\ \bibinfo {author} {\bibfnamefont {D.}~\bibnamefont
  {Rubiera-Garcia}},\ }\href {\doibase 10.1103/PhysRevD.86.044014} {\bibfield
  {journal} {\bibinfo  {journal} {Phys. Rev. D}\ }\textbf {\bibinfo {volume}
  {86}},\ \bibinfo {pages} {044014} (\bibinfo {year} {2012})},\ \Eprint
  {http://arxiv.org/abs/1207.6004} {arXiv:1207.6004 [gr-qc]} \BibitemShut
  {NoStop}%
\bibitem [{\citenamefont {Balakin}\ \emph
  {et~al.}(2016{\natexlab{a}})\citenamefont {Balakin}, \citenamefont {Lemos},\
  and\ \citenamefont {Zayats}}]{Balakin:2015gpq}%
  \BibitemOpen
  \bibfield  {author} {\bibinfo {author} {\bibfnamefont {A.~B.}\ \bibnamefont
  {Balakin}}, \bibinfo {author} {\bibfnamefont {J.~P.~S.}\ \bibnamefont
  {Lemos}}, \ and\ \bibinfo {author} {\bibfnamefont {A.~E.}\ \bibnamefont
  {Zayats}},\ }\href {\doibase 10.1103/PhysRevD.93.024008} {\bibfield
  {journal} {\bibinfo  {journal} {Phys. Rev. D}\ }\textbf {\bibinfo {volume}
  {93}},\ \bibinfo {pages} {024008} (\bibinfo {year} {2016}{\natexlab{a}})},\
  \Eprint {http://arxiv.org/abs/1512.02653} {arXiv:1512.02653 [gr-qc]}
  \BibitemShut {NoStop}%
\bibitem [{\citenamefont {Olmo}\ and\ \citenamefont
  {Rubiera-Garcia}(2015)}]{Olmo:2015axa}%
  \BibitemOpen
  \bibfield  {author} {\bibinfo {author} {\bibfnamefont {G.~J.}\ \bibnamefont
  {Olmo}}\ and\ \bibinfo {author} {\bibfnamefont {D.}~\bibnamefont
  {Rubiera-Garcia}},\ }\href {\doibase 10.3390/universe1020173} {\bibfield
  {journal} {\bibinfo  {journal} {Universe}\ }\textbf {\bibinfo {volume} {1}},\
  \bibinfo {pages} {173} (\bibinfo {year} {2015})},\ \Eprint
  {http://arxiv.org/abs/1509.02430} {arXiv:1509.02430 [hep-th]} \BibitemShut
  {NoStop}%
\bibitem [{\citenamefont {Fan}\ and\ \citenamefont {Wang}(2016)}]{Fan:2016hvf}%
  \BibitemOpen
  \bibfield  {author} {\bibinfo {author} {\bibfnamefont {Z.-Y.}\ \bibnamefont
  {Fan}}\ and\ \bibinfo {author} {\bibfnamefont {X.}~\bibnamefont {Wang}},\
  }\href {\doibase 10.1103/PhysRevD.94.124027} {\bibfield  {journal} {\bibinfo
  {journal} {Phys. Rev. D}\ }\textbf {\bibinfo {volume} {94}},\ \bibinfo
  {pages} {124027} (\bibinfo {year} {2016})},\ \Eprint
  {http://arxiv.org/abs/1610.02636} {arXiv:1610.02636 [gr-qc]} \BibitemShut
  {NoStop}%
\bibitem [{\citenamefont {Chamseddine}\ and\ \citenamefont
  {Mukhanov}(2017)}]{Chamseddine:2016ktu}%
  \BibitemOpen
  \bibfield  {author} {\bibinfo {author} {\bibfnamefont {A.~H.}\ \bibnamefont
  {Chamseddine}}\ and\ \bibinfo {author} {\bibfnamefont {V.}~\bibnamefont
  {Mukhanov}},\ }\href {\doibase 10.1140/epjc/s10052-017-4759-z} {\bibfield
  {journal} {\bibinfo  {journal} {Eur. Phys. J. C}\ }\textbf {\bibinfo {volume}
  {77}},\ \bibinfo {pages} {183} (\bibinfo {year} {2017})},\ \Eprint
  {http://arxiv.org/abs/1612.05861} {arXiv:1612.05861 [gr-qc]} \BibitemShut
  {NoStop}%
\bibitem [{\citenamefont {Menchon}\ \emph {et~al.}(2017)\citenamefont
  {Menchon}, \citenamefont {Olmo},\ and\ \citenamefont
  {Rubiera-Garcia}}]{Menchon:2017qed}%
  \BibitemOpen
  \bibfield  {author} {\bibinfo {author} {\bibfnamefont {C.}~\bibnamefont
  {Menchon}}, \bibinfo {author} {\bibfnamefont {G.~J.}\ \bibnamefont {Olmo}}, \
  and\ \bibinfo {author} {\bibfnamefont {D.}~\bibnamefont {Rubiera-Garcia}},\
  }\href {\doibase 10.1103/PhysRevD.96.104028} {\bibfield  {journal} {\bibinfo
  {journal} {Phys. Rev.}\ }\textbf {\bibinfo {volume} {D96}},\ \bibinfo {pages}
  {104028} (\bibinfo {year} {2017})},\ \Eprint
  {http://arxiv.org/abs/1709.09592} {arXiv:1709.09592 [gr-qc]} \BibitemShut
  {NoStop}%
\bibitem [{\citenamefont {Sert}(2016)}]{Sert:2015ykz}%
  \BibitemOpen
  \bibfield  {author} {\bibinfo {author} {\bibfnamefont {O.}~\bibnamefont
  {Sert}},\ }\href {\doibase 10.1063/1.4944428} {\bibfield  {journal} {\bibinfo
   {journal} {J. Math. Phys.}\ }\textbf {\bibinfo {volume} {57}},\ \bibinfo
  {pages} {032501} (\bibinfo {year} {2016})},\ \Eprint
  {http://arxiv.org/abs/1512.01172} {arXiv:1512.01172 [gr-qc]} \BibitemShut
  {NoStop}%
\bibitem [{\citenamefont {Buoninfante}\ \emph
  {et~al.}(2018{\natexlab{a}})\citenamefont {Buoninfante}, \citenamefont
  {Koshelev}, \citenamefont {Lambiase},\ and\ \citenamefont
  {Mazumdar}}]{Buoninfante:2018xiw}%
  \BibitemOpen
  \bibfield  {author} {\bibinfo {author} {\bibfnamefont {L.}~\bibnamefont
  {Buoninfante}}, \bibinfo {author} {\bibfnamefont {A.~S.}\ \bibnamefont
  {Koshelev}}, \bibinfo {author} {\bibfnamefont {G.}~\bibnamefont {Lambiase}},
  \ and\ \bibinfo {author} {\bibfnamefont {A.}~\bibnamefont {Mazumdar}},\
  }\href {\doibase 10.1088/1475-7516/2018/09/034} {\bibfield  {journal}
  {\bibinfo  {journal} {JCAP}\ }\textbf {\bibinfo {volume} {09}},\ \bibinfo
  {pages} {034} (\bibinfo {year} {2018}{\natexlab{a}})},\ \Eprint
  {http://arxiv.org/abs/1802.00399} {arXiv:1802.00399 [gr-qc]} \BibitemShut
  {NoStop}%
\bibitem [{\citenamefont {Buoninfante}\ \emph
  {et~al.}(2018{\natexlab{b}})\citenamefont {Buoninfante}, \citenamefont
  {Harmsen}, \citenamefont {Maheshwari},\ and\ \citenamefont
  {Mazumdar}}]{Buoninfante:2018stt}%
  \BibitemOpen
  \bibfield  {author} {\bibinfo {author} {\bibfnamefont {L.}~\bibnamefont
  {Buoninfante}}, \bibinfo {author} {\bibfnamefont {G.}~\bibnamefont
  {Harmsen}}, \bibinfo {author} {\bibfnamefont {S.}~\bibnamefont {Maheshwari}},
  \ and\ \bibinfo {author} {\bibfnamefont {A.}~\bibnamefont {Mazumdar}},\
  }\href {\doibase 10.1103/PhysRevD.98.084009} {\bibfield  {journal} {\bibinfo
  {journal} {Phys. Rev. D}\ }\textbf {\bibinfo {volume} {98}},\ \bibinfo
  {pages} {084009} (\bibinfo {year} {2018}{\natexlab{b}})},\ \Eprint
  {http://arxiv.org/abs/1804.09624} {arXiv:1804.09624 [gr-qc]} \BibitemShut
  {NoStop}%
\bibitem [{\citenamefont {Giacchini}\ and\ \citenamefont
  {de~Paula~Netto}(2019)}]{Giacchini:2018gxp}%
  \BibitemOpen
  \bibfield  {author} {\bibinfo {author} {\bibfnamefont {B.~L.}\ \bibnamefont
  {Giacchini}}\ and\ \bibinfo {author} {\bibfnamefont {T.}~\bibnamefont
  {de~Paula~Netto}},\ }\href {\doibase 10.1140/epjc/s10052-019-6727-2}
  {\bibfield  {journal} {\bibinfo  {journal} {Eur. Phys. J. C}\ }\textbf
  {\bibinfo {volume} {79}},\ \bibinfo {pages} {217} (\bibinfo {year} {2019})},\
  \Eprint {http://arxiv.org/abs/1806.05664} {arXiv:1806.05664 [gr-qc]}
  \BibitemShut {NoStop}%
\bibitem [{\citenamefont {Cano}\ \emph {et~al.}(2019)\citenamefont {Cano},
  \citenamefont {Chimento}, \citenamefont {Ort\'in},\ and\ \citenamefont
  {Ruip\'erez}}]{Cano:2018aod}%
  \BibitemOpen
  \bibfield  {author} {\bibinfo {author} {\bibfnamefont {P.~A.}\ \bibnamefont
  {Cano}}, \bibinfo {author} {\bibfnamefont {S.}~\bibnamefont {Chimento}},
  \bibinfo {author} {\bibfnamefont {T.}~\bibnamefont {Ort\'in}}, \ and\
  \bibinfo {author} {\bibfnamefont {A.}~\bibnamefont {Ruip\'erez}},\ }\href
  {\doibase 10.1103/PhysRevD.99.046014} {\bibfield  {journal} {\bibinfo
  {journal} {Phys. Rev. D}\ }\textbf {\bibinfo {volume} {99}},\ \bibinfo
  {pages} {046014} (\bibinfo {year} {2019})},\ \Eprint
  {http://arxiv.org/abs/1806.08377} {arXiv:1806.08377 [hep-th]} \BibitemShut
  {NoStop}%
\bibitem [{Note1()}]{Note1}%
  \BibitemOpen
  \bibinfo {note} {In the case of magnetically-charged solutions, regular black
  holes were studied in Ref.~\cite {Balakin:2015gpq} in a similar set-up. The
  magnetic field has a singularity though. On the other hand, in the solutions
  of Refs.~\cite {AyonBeato:1998ub,AyonBeato:1999ec,AyonBeato:1999rg} the
  charge and mass are not free parameters.}\BibitemShut {Stop}%
\bibitem [{Note2()}]{Note2}%
  \BibitemOpen
  \bibinfo {note} {The generalized Kronecker delta is defined as $\protect
  \tensor {\delta }{^{\mu \nu }_{\rho \sigma }}=\protect \tensor {\delta
  }{^{[\mu }_{[\rho }}\protect \tensor {\delta }{^{\nu ]}_{\sigma
  ]}}$}\BibitemShut {NoStop}%
\bibitem [{Note3()}]{Note3}%
  \BibitemOpen
  \bibinfo {note} {This follows from the fact that $\protect \tensor {(\chi
  ^{-1})}{^{\mu \nu }_{\rho \sigma }}=6\protect \tensor {\delta }{^{[\mu \nu
  }_{\rho \sigma }}\protect \tensor {\protect \mathcal {Q}}{^{\alpha \beta
  ]}_{\alpha \beta }}\protect \tmspace +\thinmuskip {.1667em}$}\BibitemShut
  {NoStop}%
\bibitem [{\citenamefont {Bueno}\ and\ \citenamefont
  {Cano}(2016{\natexlab{a}})}]{PabloPablo}%
  \BibitemOpen
  \bibfield  {author} {\bibinfo {author} {\bibfnamefont {P.}~\bibnamefont
  {Bueno}}\ and\ \bibinfo {author} {\bibfnamefont {P.~A.}\ \bibnamefont
  {Cano}},\ }\href {\doibase 10.1103/PhysRevD.94.104005} {\bibfield  {journal}
  {\bibinfo  {journal} {Phys. Rev.}\ }\textbf {\bibinfo {volume} {D94}},\
  \bibinfo {pages} {104005} (\bibinfo {year} {2016}{\natexlab{a}})},\ \Eprint
  {http://arxiv.org/abs/1607.06463} {arXiv:1607.06463 [hep-th]} \BibitemShut
  {NoStop}%
\bibitem [{\citenamefont {Hennigar}\ and\ \citenamefont
  {Mann}(2017)}]{Hennigar:2016gkm}%
  \BibitemOpen
  \bibfield  {author} {\bibinfo {author} {\bibfnamefont {R.~A.}\ \bibnamefont
  {Hennigar}}\ and\ \bibinfo {author} {\bibfnamefont {R.~B.}\ \bibnamefont
  {Mann}},\ }\href {\doibase 10.1103/PhysRevD.95.064055} {\bibfield  {journal}
  {\bibinfo  {journal} {Phys. Rev.}\ }\textbf {\bibinfo {volume} {D95}},\
  \bibinfo {pages} {064055} (\bibinfo {year} {2017})},\ \Eprint
  {http://arxiv.org/abs/1610.06675} {arXiv:1610.06675 [hep-th]} \BibitemShut
  {NoStop}%
\bibitem [{\citenamefont {Bueno}\ and\ \citenamefont
  {Cano}(2016{\natexlab{b}})}]{PabloPablo2}%
  \BibitemOpen
  \bibfield  {author} {\bibinfo {author} {\bibfnamefont {P.}~\bibnamefont
  {Bueno}}\ and\ \bibinfo {author} {\bibfnamefont {P.~A.}\ \bibnamefont
  {Cano}},\ }\href {\doibase 10.1103/PhysRevD.94.124051} {\bibfield  {journal}
  {\bibinfo  {journal} {Phys. Rev.}\ }\textbf {\bibinfo {volume} {D94}},\
  \bibinfo {pages} {124051} (\bibinfo {year} {2016}{\natexlab{b}})},\ \Eprint
  {http://arxiv.org/abs/1610.08019} {arXiv:1610.08019 [hep-th]} \BibitemShut
  {NoStop}%
\bibitem [{\citenamefont {Hennigar}\ \emph {et~al.}(2017)\citenamefont
  {Hennigar}, \citenamefont {Kubiznak},\ and\ \citenamefont
  {Mann}}]{Hennigar:2017ego}%
  \BibitemOpen
  \bibfield  {author} {\bibinfo {author} {\bibfnamefont {R.~A.}\ \bibnamefont
  {Hennigar}}, \bibinfo {author} {\bibfnamefont {D.}~\bibnamefont {Kubiznak}},
  \ and\ \bibinfo {author} {\bibfnamefont {R.~B.}\ \bibnamefont {Mann}},\
  }\href {\doibase 10.1103/PhysRevD.95.104042} {\bibfield  {journal} {\bibinfo
  {journal} {Phys. Rev.}\ }\textbf {\bibinfo {volume} {D95}},\ \bibinfo {pages}
  {104042} (\bibinfo {year} {2017})},\ \Eprint
  {http://arxiv.org/abs/1703.01631} {arXiv:1703.01631 [hep-th]} \BibitemShut
  {NoStop}%
\bibitem [{\citenamefont {Bueno}\ and\ \citenamefont
  {Cano}(2017)}]{PabloPablo3}%
  \BibitemOpen
  \bibfield  {author} {\bibinfo {author} {\bibfnamefont {P.}~\bibnamefont
  {Bueno}}\ and\ \bibinfo {author} {\bibfnamefont {P.~A.}\ \bibnamefont
  {Cano}},\ }\href {\doibase 10.1088/1361-6382/aa8056} {\bibfield  {journal}
  {\bibinfo  {journal} {Class. Quant. Grav.}\ }\textbf {\bibinfo {volume}
  {34}},\ \bibinfo {pages} {175008} (\bibinfo {year} {2017})},\ \Eprint
  {http://arxiv.org/abs/1703.04625} {arXiv:1703.04625 [hep-th]} \BibitemShut
  {NoStop}%
\bibitem [{\citenamefont {Cano}\ and\ \citenamefont
  {Murcia}(2020)}]{Cano:2020qhy}%
  \BibitemOpen
  \bibfield  {author} {\bibinfo {author} {\bibfnamefont {P.~A.}\ \bibnamefont
  {Cano}}\ and\ \bibinfo {author} {\bibfnamefont {A.}~\bibnamefont {Murcia}},\
  }\href@noop {} {\  (\bibinfo {year} {2020})},\ \Eprint
  {http://arxiv.org/abs/2007.04331} {arXiv:2007.04331 [hep-th]} \BibitemShut
  {NoStop}%
\bibitem [{\citenamefont {Mueller-Hoissen}\ and\ \citenamefont
  {Sippel}(1988)}]{MuellerHoissen:1988bp}%
  \BibitemOpen
  \bibfield  {author} {\bibinfo {author} {\bibfnamefont {F.}~\bibnamefont
  {Mueller-Hoissen}}\ and\ \bibinfo {author} {\bibfnamefont {R.}~\bibnamefont
  {Sippel}},\ }\href {\doibase 10.1088/0264-9381/5/11/010} {\bibfield
  {journal} {\bibinfo  {journal} {Class. Quant. Grav.}\ }\textbf {\bibinfo
  {volume} {5}},\ \bibinfo {pages} {1473} (\bibinfo {year} {1988})}\BibitemShut
  {NoStop}%
\bibitem [{\citenamefont {Balakin}\ \emph
  {et~al.}(2016{\natexlab{b}})\citenamefont {Balakin}, \citenamefont {Lemos},\
  and\ \citenamefont {Zayats}}]{Balakin:2016mnn}%
  \BibitemOpen
  \bibfield  {author} {\bibinfo {author} {\bibfnamefont {A.~B.}\ \bibnamefont
  {Balakin}}, \bibinfo {author} {\bibfnamefont {J.~P.~S.}\ \bibnamefont
  {Lemos}}, \ and\ \bibinfo {author} {\bibfnamefont {A.~E.}\ \bibnamefont
  {Zayats}},\ }\href {\doibase 10.1103/PhysRevD.93.084004} {\bibfield
  {journal} {\bibinfo  {journal} {Phys. Rev. D}\ }\textbf {\bibinfo {volume}
  {93}},\ \bibinfo {pages} {084004} (\bibinfo {year} {2016}{\natexlab{b}})},\
  \Eprint {http://arxiv.org/abs/1603.02676} {arXiv:1603.02676 [gr-qc]}
  \BibitemShut {NoStop}%
\end{thebibliography}%

\end{document}